\documentclass[conference]{IEEEtran}
\ifCLASSINFOpdf
\else
\fi

\usepackage{graphicx,color}				
\usepackage{amssymb,amsmath,amsfonts,amsthm}  % Math packages

\usepackage{epstopdf}
\usepackage{lipsum} 

\usepackage{algorithm,caption}
\usepackage[noend]{algpseudocode}

\newcounter{MYtempeqncnt}

\def \tr {\text{tr}}

\begin{document}
%\title{FEC Code Anchored MIMO Receiver\\ Using Semi-definite Relaxation}
\title{Iterative Turbo Receiver for LDPC-Coded MIMO Systems Based on Semi-definite Relaxation}

% =========== for conference ==============
\author{\IEEEauthorblockN{Kun Wang}
\IEEEauthorblockA{Qualcomm Technologies, Inc.\\
3165 Kifer Road\\
Santa Clara, CA 95051, USA\\
Email: kunwang@ieee.org}
\and
\IEEEauthorblockN{Zhi Ding}
\IEEEauthorblockA{Dept. of Electrical and Computer Engineering\\
University of California, Davis\\
Davis, CA 95616, USA\\
Email: zding@ucdavis.edu}
}
\maketitle

\begin{abstract}
In this work, we develop a new iterative turbo receiver for 
LDPC-coded multi-antenna systems based on semi-definite relaxation (SDR). 
For a classical turbo receiver, forward error correction (FEC) code is only used at decoder.
Nonetheless, by taking advantage of FEC code in the detection stage,
our proposed SDR detector can output extrinsic information with much improved reliability 
to the decoder. 
We also propose a simplified SDR turbo receiver that solves only one SDR problem per codeword 
instead of solving multiple SDR problems in the iterative turbo processing. 
This scheme significantly reduces the time complexity of SDR turbo receiver,
while the error performance remains similar as before. 
In fact, our simplified scheme is generic in the sense that 
it is applicable to any list-based iterative receivers.
\end{abstract}


% \fi
\IEEEpeerreviewmaketitle

\section{Introduction}  \label{sec:intro}

% MIMO advances
Multiple-input multiple-output (MIMO) technology offers the potential for
high data rates and reliable transmissions,
where the underlying premise is advanced design of wireless transceiver.  
In the receiver end, turbo processing is known to be capable of 
approaching MIMO capacity by exchanging extrinsic information 
between detector and decoder \cite{hochwald2003achieving}.  
In spite of the near-capacity capability, the soft detector in the turbo receiver 
incurs exponential complexity in the computation of exact extrinsic information, 
which is often in the format of log-likelihood ratio (LLR). 
Therefore, it has stimulated a wide interest to reduce the complexity of turbo receiver,
possibly with a tolerable performance degradation.

% Different turbo approaches
To lower the complexity of exact LLR computation, max-log approximation is often used.
Nonetheless, it is still NP-hard after this approximation. 
Tree search methods were then proposed to find the optimal solutions,
whose computation costs however remain exponential in terms of 
both worst-case and average complexity \cite{jalden2005complexity}.
Further, a number of reduced tree search approaches were developed
to produce relatively good suboptimal solutions \cite{studer2010soft}.
In a recent decade or so, SDR has become a popular technique 
to approximate the maximum-likelihood (ML) solutions because of 
its upper-bounded polynomial complexity and its guaranteed approximation error \cite{luo2010semidefinite}.
SDR has also been applied to the design of lower-complexity turbo receiver.
Instead of enumerating through the exponential-sized candidate list,
the authors of \cite{steingrimsson2003soft} solve one SDR problem for each coded bit 
and this approach results in no performance loss. 
The authors of \cite{nekuii2011efficient} further developed two
soft-output SDR detectors that are significantly less complex 
while suffering slight degradation than full-list turbo receivers in performance.
More recently, as a follow-up paper of \cite{nekuii2011efficient}, the 
authors of \cite{salmani2017semidefinite} extended the 
efficient SDR receivers from 4-QAM (QPSK) to higher-order QAM signaling
by presenting two customized algorithms for solving the SDR demodulators.

% Our contribution
In this work, we present a new SDR-based turbo receiver for LDPC-coded MIMO systems.
In our detector design, FEC codes not only are used for decoding, but also are
integrated as constraints within the detection optimization formulation
\cite{wang2015diversity,wang2016robust,wang2014joint,wang2018non}. 
The proposed soft-in soft-out joint SDR detector demonstrates substantial performance
gain through iterative turbo processing. 
The joint SDR has significantly lower complexity compared with the original full-list detector, 
while achieving similar bit error rate (BER) in overall performance. 
Furthermore, we also present a simplified joint SDR turbo receiver. 
In this new approach, only one SDR is solved in the initial iteration
for each codeword, unlike the works that require multiple SDR solutions. 
In subsequent iterations, we propose a simple approximation to
generate the requisite output extrinsic information for turbo message passing. 
Compared with existing SDR-based turbo receivers, 
both the receiver in \cite{steingrimsson2003soft} and our new
work retain the original turbo detection performance, 
but the complexity of our proposed scheme is lower because we solve fewer SDR problems per codeword.  
Moreover, the receivers presented in \cite{nekuii2011efficient} used 
the randomization approach or Bernoulli trials to generate a preliminary candidate list
and then enriched the list by bit flipping. 
However, based on our reliable joint SDR solution, 
we can directly generate the candidate list without additional steps. 
Furthermore, the methods in \cite{nekuii2011efficient} slightly trade BER performance for 
low complexity, as shown in the simulations.

\section{System Model}  \label{sec:sys_model}

\subsection{MIMO System Model}
% Complex model 
We consider an $N_t$-input $N_r$-output spatial multiplexing MIMO system.
The channel is assumed to be flat-fading.
The baseband equivalent model at time $k$ can be expressed as
\begin{equation} \label{eq:mimo_complex}
\mathbf{y}_k^c = \mathbf{H}_k^c \mathbf{s}_k^c + \mathbf{n}_k^c, \quad k = 1, \ldots, K,
\end{equation}
where $\mathbf{y}_k^c \in \mathbb{C}^{N_r \times 1}$ is the received signal,
$\mathbf{H}_k^c \in \mathbb{C}^{N_r \times N_t}$ denotes the MIMO channel matrix,
$\mathbf{s}_k^c \in \mathbb{C}^{N_t \times 1}$ is the transmitted signal, and 
$\mathbf{n}_k^c  \in \mathbb{C}^{N_r \times 1}$ is an additive Gaussian noise vector, 
each element of which is independent and follows $\mathcal{CN}(0, 2 \sigma_n^2)$. 

% Real model
To simplify subsequent problem formulation, the complex-valued
signal model can be transformed into the real field by letting
\begin{equation*}
\mathbf{y}_k = 
\begin{bmatrix}
\text{Re}\{ \mathbf{y}_k^c  \} \\
\text{Im}\{ \mathbf{y}_k^c  \}
\end{bmatrix},
\mathbf{s}_k = 
\begin{bmatrix}
\text{Re}\{ \mathbf{s}_k^c  \} \\
\text{Im}\{ \mathbf{s}_k^c  \}
\end{bmatrix},
\mathbf{n}_k = 
\begin{bmatrix}
\text{Re}\{ \mathbf{n}_k^c  \} \\
\text{Im}\{ \mathbf{n}_k^c  \}
\end{bmatrix},
\end{equation*}
and
\begin{equation*}
\mathbf{H}_k = 
\begin{bmatrix}
\text{Re}\{ \mathbf{H}_k^c  \}  & - \text{Im}\{ \mathbf{H}_k^c  \}\\
\text{Im}\{ \mathbf{H}_k^c  \} & \text{Re}\{ \mathbf{H}_k^c  \}
\end{bmatrix}.
\end{equation*}
Consequently, the transmission equation is given by
\begin{equation} \label{eq:mimo_real}
\mathbf{y}_k = \mathbf{H}_k \mathbf{s}_k + \mathbf{n}_k, \quad k = 1, \ldots, K.
\end{equation}

In this study, we choose capacity-approaching LDPC code for the purpose of forward error correction. 
Further, we assume the transmitted symbols are generated from QPSK constellation,
i.e., $s_{k,i}^c \in \{ \pm1 \pm j \}$ for $k = 1, \ldots, K$ and $i = 1, \ldots, N_t$.
The spatial multiplexing is done by placing the codeword
first along the spatial dimension and then along the temporal dimension.

\subsection{Turbo Receiver Structure}
\begin{figure}[!tb]
\centering
\centerline{\includegraphics[width=9cm]{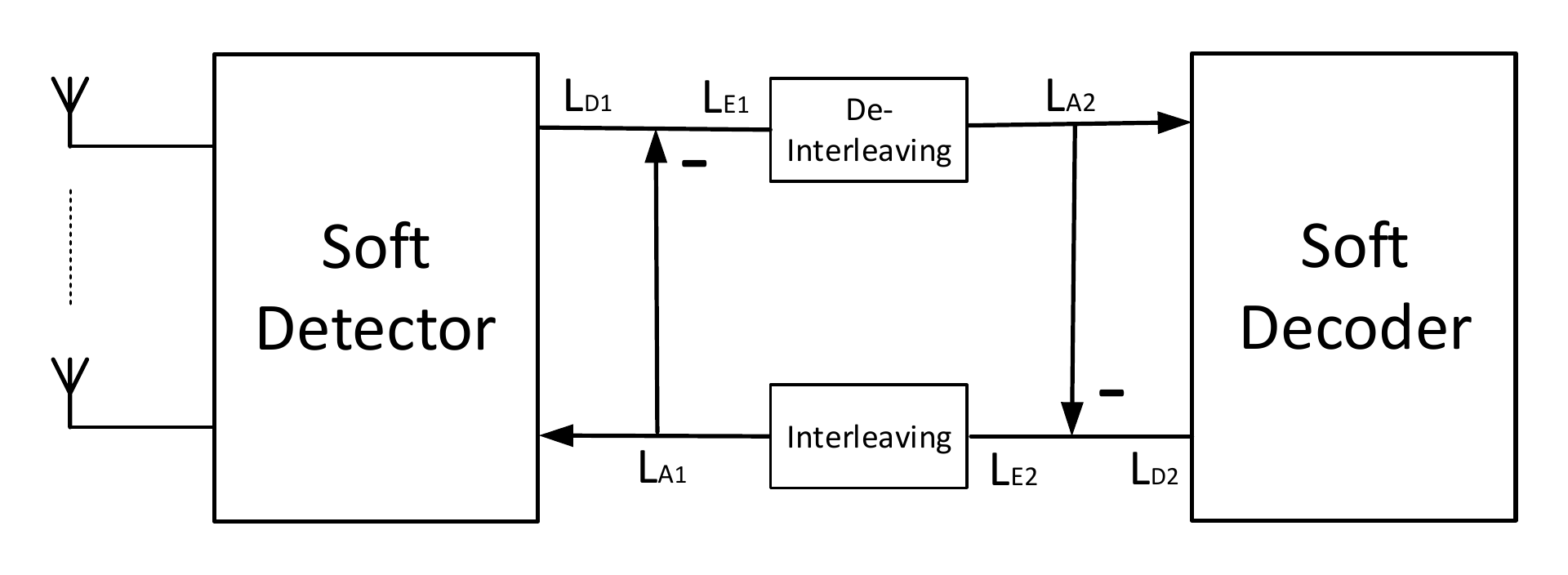}}
\caption{\small{Structure of Turbo Receiver.}}
\label{fig:turbo_receiver}
\vspace{-3mm}
\end{figure}

% structure figure
Given the system model above,
the structure of a typical turbo receiver for MIMO systems is shown in Fig.~\ref{fig:turbo_receiver}.
The major blocks include a MIMO detector and a channel decoder,
with extrinsic information exchanging between them.
Note that both detector and decoder are soft-in and soft-out. 

% structure details
Specifically, the MIMO detector takes in received signals 
and $\textit{a priori}$ information (often in the format of LLR),
and outputs soft information of each bit,
denoted by $L_{D1}$ in the figure.
After subtracting the prior information $L_{A1}$ from $L_{D1}$,
the extrinsic information is given by $L_{E1} = L_{D1} - L_{A1}$.
Then $L_{E1}$ is de-interleaved to become $L_{A2}$ as the input to channel decoder.
The path from decoder to detector follows similar processing. 
For LDPC decoding, sum-product algorithm (SPA) is often used
because of its superior performance and relatively low complexity. 
In this work, we use the ``standard'' log-domain SPA decoder.
Thus, our design focus is the soft MIMO detector.  

% List-based LLR Generation
Before diving into the detector design, we review the classical approach of list-based LLR generation.
Let $\mathbf{s}_k = \mathcal{M}(\mathbf{b}_k)$ denote
the QPSK modulator applied to a vector of polarized bits ($\pm1$),
and $\mathbf{L}_{A1,k}$ is the prior LLR vector corresponding to $\mathbf{b}_k$.
Here, we note that the polarized bit $b_{i,k} = 1 - 2 c_{i,k}$ for coded bit $c_{i,k} \in \{0,1\}$,
where subscript $(i,k)$ denotes the $i$-th bit at time $k$.
Further, let the vector with superscript $[i]$ denote a vector excluding the $i$-th element.
Also, denote $\mathcal{L} = \{-1,+1\}^{2N_t}$ 
and $\mathcal{L}_{i,\pm1} = \left\{ \mathbf{b} \in \mathcal{L} \, | \, b_i = \pm 1 \right\}$. 
Following the derivations in \cite{hochwald2003achieving}, 
the extrinsic LLR of bit $b_{i,k}$ with max-log approximation is given by
\begin{equation} \label{eq:extr_llr}
\begin{split}
L_{E1} (b_{i,k}) \approx
\max_{\mathbf{b}_k \in \mathcal{L}_{i,+1}} \left\{ -\frac{|| \mathbf{y}_k - \mathbf{H}_k \mathbf{s}_k ||^2}{2 \sigma_n^2} + 
\frac{(\mathbf{L}_{A1,k}^{[i]})^T \mathbf{b}_{k}^{[i]}}{2} \right\} \\
- \max_{\mathbf{b}_k \in \mathcal{L}_{i,-1}} \left\{ -\frac{|| \mathbf{y}_k - \mathbf{H}_k \mathbf{s}_k ||^2}{2 \sigma_n^2} + 
\frac{(\mathbf{L}_{A1,k}^{[i]})^T \mathbf{b}_{k}^{[i]}}{2} \right\} 
\end{split}
\end{equation}

It is noted that the cardinality of $\mathcal{L}$ is exponential in $N_t$.
More specifically, in the case of QPSK,  $|\mathcal{L}| = 4^{N_t}$.
Thus, it is imperative to reduce the list size for practical use, 
especially in the coming era of massive MIMO.
On the other hand, to avoid severe LLR quality degradation,
 the reduced list should contain the true maximizer 
or at least the candidates that are close to the true maximizer.

\section{Iterative Turbo SDR Receiver}  \label{sec:turbo_sdr}

% ############ Double-column equations ###############
% Reference: Page 12 
% http://ctan.math.washington.edu/tex-archive/macros/latex/contrib/IEEEtran/IEEEtran_HOWTO.pdf
\begin{figure*}[!t]
% ensure that we have normalsize text
\normalsize
% Store the current equation number.
\setcounter{MYtempeqncnt}{\value{equation}}
% Set the equation number to one less than the one
% desired for the first equation here.
% The value here will have to changed if equations
% are added or removed prior to the place these
% equations are referenced in the main text.
\setcounter{equation}{11}
\begin{equation} \label{eq:joint_map_sdr}
\begin{aligned}
& \underset{\{\mathbf{X}_k, f_n\}}{\text{min.}}
& &  \sum_{k=1}^K \tr(\mathbf{C}_k \mathbf{X}_k) + 2 \sigma_n^2 \mathbf{L}_{A1}^T \mathbf{f} \\
& \text{s.t.}
& & \mathbf{X}_k(i,i) = 1, \, \mathbf{X}_k \succeq 0, \quad k = 1, \ldots, K, i = 1, \ldots, 2N_t+1, \\
&
& & \mathbf{X}_k(i,2N_t+1) = 1 - 2 f_{2N_t(k-1)+2i-1}, \quad k = 1, \ldots, K, i = 1, \ldots, N_t, \\
&
& & \mathbf{X}_k(i+N_t,2N_t+1) = 1 - 2 f_{2N_t(k-1)+2i}, \quad k = 1, \ldots, K, i = 1, \ldots, N_t, \\
&
& & \sum_{ n \in \mathcal{F} } f_n - \sum_{ n \in \mathcal{N}_m \backslash \mathcal{F}} f_n \leq |\mathcal{F}| - 1, \quad \forall m \in \mathcal{M}, \forall \mathcal{F} \in \mathcal{S}; \\
&&& 0  \leq f_n \leq 1, \quad \forall n \in \mathcal{N}.
\end{aligned}
\end{equation}
% Restore the current equation number.
\setcounter{equation}{\value{MYtempeqncnt}}
% IEEE uses as a separator
\hrulefill
% The spacer can be tweaked to stop underfull vboxes.
\vspace*{4pt}
\end{figure*}
% #########################################

\subsection{Non-iterative Joint SDR Detection}
Based on the assumption of Gaussian noise, it can be easily shown that the optimal ML detection
is equivalent to the following discrete least squares problem
\begin{equation} \label{eq:ml_detection}
\underset{\mathbf{x}_k \in \{ \pm 1 \}^{2N_t}}{\text{min.}} \;
\sum_{k=1}^K \Vert \mathbf{y}_k - \mathbf{H}_k \mathbf{x}_k \Vert^2.
\end{equation}

However, this problem is NP-hard. 
Instead, SDR can generate an \textit{approximate} solution to the ML problem
in polynomial time.
To solve it via SDR, define the rank-1 semi-definite matrix
\begin{equation} \label{eq:rank1_matrix}
\mathbf{X}_k = 
\begin{bmatrix}
\mathbf{x}_k \\
t_k
\end{bmatrix}
\begin{bmatrix}
\mathbf{x}_k^T & t_k
\end{bmatrix}
=
\begin{bmatrix}
\mathbf{x}_k \mathbf{x}_k^T & t_k \mathbf{x}_k \\
t_k \mathbf{x}_k^T & t_k^2
\end{bmatrix},
\end{equation}
and denote the cost matrix by
\begin{equation} \label{eq:cost_matrix}
\mathbf{C}_k = 
\begin{bmatrix}
\mathbf{H}_k^T \mathbf{H}_k & \mathbf{H}_k^T \mathbf{y}_k  \\
-\mathbf{y}_k^T \mathbf{H}_k & || \mathbf{y}_k ||^2
\end{bmatrix}.
\end{equation}
Using the property of trace $\mathbf{v}^T\mathbf{Q}\mathbf{v} = \tr(\mathbf{v}^T\mathbf{Q}\mathbf{v}) = \tr(\mathbf{Q}\mathbf{v}\mathbf{v}^T)$,
ML detection in Eq.~(\ref{eq:ml_detection}) 
can be relaxed to SDR by removing the rank-1 constraint on $\mathbf{X}_k$.
The SDR formulation is therefore
\begin{equation} \label{eq:disjoint_sdr}
\begin{aligned}
& \underset{\{\mathbf{X}_k\}}{\text{min.}}
& &  \sum_{k=1}^K \tr(\mathbf{C}_k \mathbf{X}_k) \\
& \text{s.t.}
& & \mathbf{X}_k(i,i) = 1, \; k = 1, \ldots, K, i = 1, \ldots, 2N_t + 1, \\
& 
& & \mathbf{X}_k \succeq 0, \; k = 1, \ldots, K.
\end{aligned}
\end{equation}

We can further enhance the SDR performance by incorporating 
LDPC code constraints, which are captured by 
the following forbidden set (FS) constraints \cite{feldman2005using}
\begin{equation} \label{eq:parity_ineq}
\sum_{ n \in \mathcal{F} } f_n - \sum_{ n \in \mathcal{N}_m \backslash \mathcal{F}} f_n \leq |\mathcal{F}| - 1, \; \forall m \in \mathcal{M},
\forall \mathcal{F} \in \mathcal{S}
\end{equation}
plus the box constraints for bit variables
\begin{equation} \label{eq:box_ineq}
 0  \leq f_n \leq 1, \quad \forall n \in \mathcal{N}.
\end{equation}

To connect the code constraints with SDR formulation,
we recognize the bit-to-symbol mapping 
for time $k = 1, \ldots, K$ and bit index $i = 1, \ldots, N_t$
simply as follows
\begin{equation} \label{eq:qpsk_gray}
\begin{split}
& \mathbf{X}_k(i,2N_t+1) = 1 - 2 f_{2N_t(k-1)+2i-1}, \\
& \mathbf{X}_k(i+N_t,2N_t+1) = 1 - 2 f_{2N_t(k-1)+2i}.
\end{split}
\end{equation} 
For the details of LDPC-integrated SDR formulation,
we refer the readers to the paper \cite{wang2018non}.

\subsection{Joint MAP-SDR Turbo Receiver}

When \textit{a priori} information of each bit is available, \textit{maximum a posterior} (MAP) criterion 
can be employed instead of ML. According to \cite{hagenauer1997turbo}, the likelihood probability
$p(\mathbf{y}_k | \mathbf{s}_k) \propto \exp(-|| \mathbf{y}_k - \mathbf{H}_k \mathbf{s}_k ||^2/(2 \sigma_n^2))$ 
and \textit{a priori} probability
$p\left(\mathbf{s}_k = \mathcal{M}(\mathbf{b}_k)\right) \propto \exp(\mathbf{L}_{A1,k}^T \mathbf{b}_k / 2)$.
Therefore, the \textit{a posterior} probability can be given as
\begin{equation} \label{eq:map_prob}
\begin{split}
p(\mathbf{s}_k | \mathbf{y}_k) 
& \propto p(\mathbf{y}_k | \mathbf{s}_k) p(\mathbf{s}_k) \\
& \propto \exp\left(-\frac{|| \mathbf{y}_k - \mathbf{H}_k \mathbf{s}_k ||^2}{2 \sigma_n^2} 
   + \frac{\mathbf{L}_{A1,k}^T \mathbf{b}_k}{2}\right).
\end{split}
\end{equation} 
After taking logarithm and summing over the $K$ time instants, MAP is equivalent to minimizing the new cost function
\begin{equation*} \label{eq:map_cost_func}
\sum_{k=1}^K \tr(\mathbf{C}_k \mathbf{X}_k) - \sigma_n^2 \mathbf{L}_{A1}^T (\mathbf{1} - 2 \mathbf{f})
 = \sum_{k=1}^K \tr(\mathbf{C}_k \mathbf{X}_k) + 2 \sigma_n^2 \mathbf{L}_{A1}^T \mathbf{f}.
 \end{equation*}
 
By integrating the constraints from Eq.~(\ref{eq:disjoint_sdr}), (\ref{eq:parity_ineq}), 
(\ref{eq:box_ineq}) and (\ref{eq:qpsk_gray}),
the optimization problem in Eq.~(\ref{eq:joint_map_sdr}) describes the new joint MAP-SDR detector. 
Notice that our MAP cost function in Eq.~(\ref{eq:joint_map_sdr}) is 
generally applicable to any QAM constellations,
whereas the approach in \cite{salmani2017semidefinite} 
was to approximate the cost function for higher order QAM.
For higher order QAM beyond QPSK, the necessary changes 
for our joint SDR receiver include box relaxation of diagonal elements of  $\mathbf{X}_k$ \cite{ma2009equivalence}
and the modification of symbol-to-bit mapping constraints.
We refer interested readers to the works \cite{wang2016fec,wang2015joint,wang2016diversity,wang2017galois,wang2018integrated} for details of 
higher order QAM mapping constraints.

%\begin{equation} \label{eq:joint_map_sdr}
%\begin{aligned}
%& \underset{\{\mathbf{X}_k, f_n\}}{\text{min.}}
%& &  \sum_{k=1}^K \tr(\mathbf{C}_k \mathbf{X}_k) + 2 \sigma_n^2 \mathbf{L}_{A1}^T \mathbf{f} \\
%& \text{s.t.}
%& & \mathbf{X}_k(i,i) = 1, \, \mathbf{X}_k \succeq 0, \quad k = 1, \ldots, K, i = 1, \ldots, 2N_t+1, \\
%&
%& & \mathbf{X}_k(i,2N_t+1) = 1 - 2 f_{2N_t(k-1)+2i-1}, \quad k = 1, \ldots, K, i = 1, \ldots, N_t, \\
%&
%& & \mathbf{X}_k(i+N_t,2N_t+1) = 1 - 2 f_{2N_t(k-1)+2i}, \quad k = 1, \ldots, K, i = 1, \ldots, N_t, \\
%&
%& & \sum_{ n \in \mathcal{F} } f_n - \sum_{ n \in \mathcal{N}_m \backslash \mathcal{F}} f_n \leq |\mathcal{F}| - 1, \quad \forall m \in \mathcal{M}, \forall \mathcal{F} \in \mathcal{S}; \\
%&&& 0  \leq f_n \leq 1, \quad \forall n \in \mathcal{N}.
%\end{aligned}
%\end{equation}

With the solution from joint MAP-SDR detector, it is unnecessary to enumerate 
over the full list $\mathcal{L}$ to generate LLRs as shown in Eq.~(\ref{eq:extr_llr}).
Instead, we can construct a subset $\mathcal{\overline{L}}_k \subseteq \mathcal{L}$, containing the 
probable candidates that are within a certain Hamming distance from the SDR optimal solution $\mathbf{b}_k^*$ \cite{love2005space}.
More specifically, $\mathcal{\overline{L}}_k = 
\left\{ \mathbf{b}'_k \in \mathcal{L} \, | \, d(\mathbf{b}'_k, \mathbf{b}_k^*) 
\leq P  \right\}$,
where the Hamming distance $d(\mathbf{b}', \mathbf{b}'') = \text{card}( \{i \, | \, b'_i \neq b''_i \})$. 
Correspondingly, we have $\mathcal{\overline{L}}_{i,k,\pm1} = \left\{ \mathbf{b}_k \in \mathcal{\overline{L}}_k \, | \, b_{i,k} = \pm 1 \right\}$.
The radius $P$ determines the cardinality of 
$\mathcal{\overline{L}}_k$, 
that is, $|\mathcal{\overline{L}}_k| = 
\sum_{j=0}^P \binom{2N_t}{j}$.
Compared to the full list's size $4^{N_t}$, this could significantly reduce the list size with the selection of a small $P$.
We now briefly summarize the steps of this novel turbo receiver: 
\begin{description}
\item[S0] To initialize, let the first iteration $\mathbf{L}_{A1} = \mathbf{0}$, and select a value $P$. 
\item[S1] Solve the joint MAP-SDR given in Eq.~(\ref{eq:joint_map_sdr}).
\item[S2] Generate a list $\mathcal{\overline{L}}_k$ with a given $P$, and generate extrinsic LLRs $\mathbf{L}_{E1}$ via Eq.~(\ref{eq:extr_llr}) with $\mathcal{L}_{i,\pm1}$ being replaced by $\mathcal{\overline{L}}_{i,k,\pm1}$.
\item[S3] Send de-interleaved $\mathbf{L}_{A2}$ to SPA decoder. If maximum iterations are reached or if all FEC parity checks are satisfied after decoding, 
stop the turbo process; Otherwise, return to S1.
\end{description}

\subsection{Simplified Turbo SDR Receiver}

One can clearly see that it is costly for our proposed turbo SDR
algorithm to solve one joint MAP-SDR in each iteration (in step S1).
To reduce receiver complexity, 
we can solve one joint MAP-SDR in the first iteration
and generate the candidate list by other means in subsequent iterations
without repeatedly solving the joint MAP-SDR. 
In fact, the authors \cite{nekuii2011efficient} proposed 
a Bernoulli randomization method to generate such a candidate list 
based only on the first iteration SDR output and subsequent decoder feedback.
We now propose another list generation method for our receiver
that is more efficient.

\begin{figure}[!tb]
	    \begin{minipage}[b]{0.48\linewidth}
	      \centering
	      \centerline{\includegraphics[width=3cm]{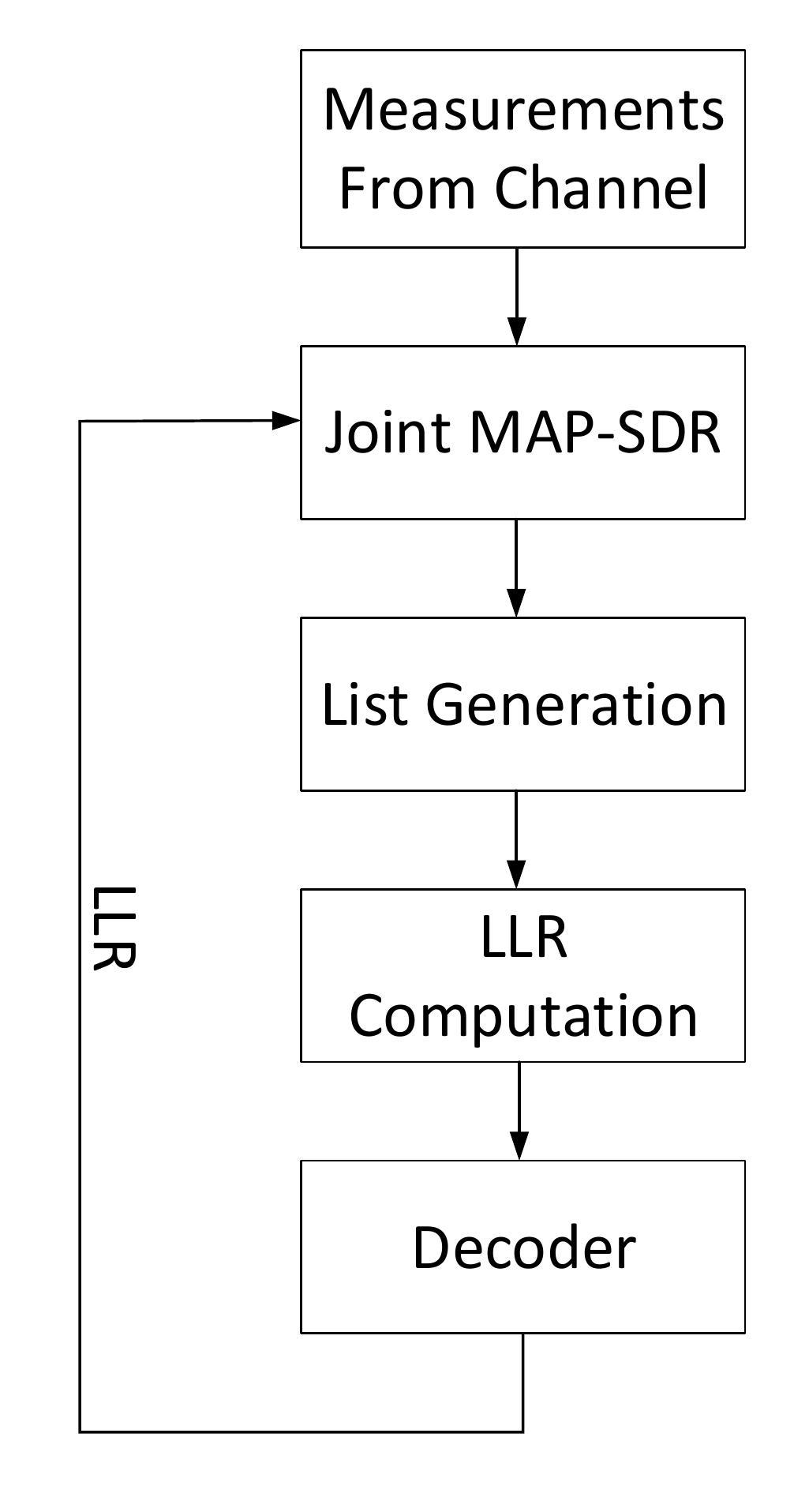}}
	      \centerline{(a)}\medskip
	    \end{minipage}
	    \hfill
	    \begin{minipage}[b]{.48\linewidth}
	      \centering
	      \centerline{\includegraphics[width=3cm]{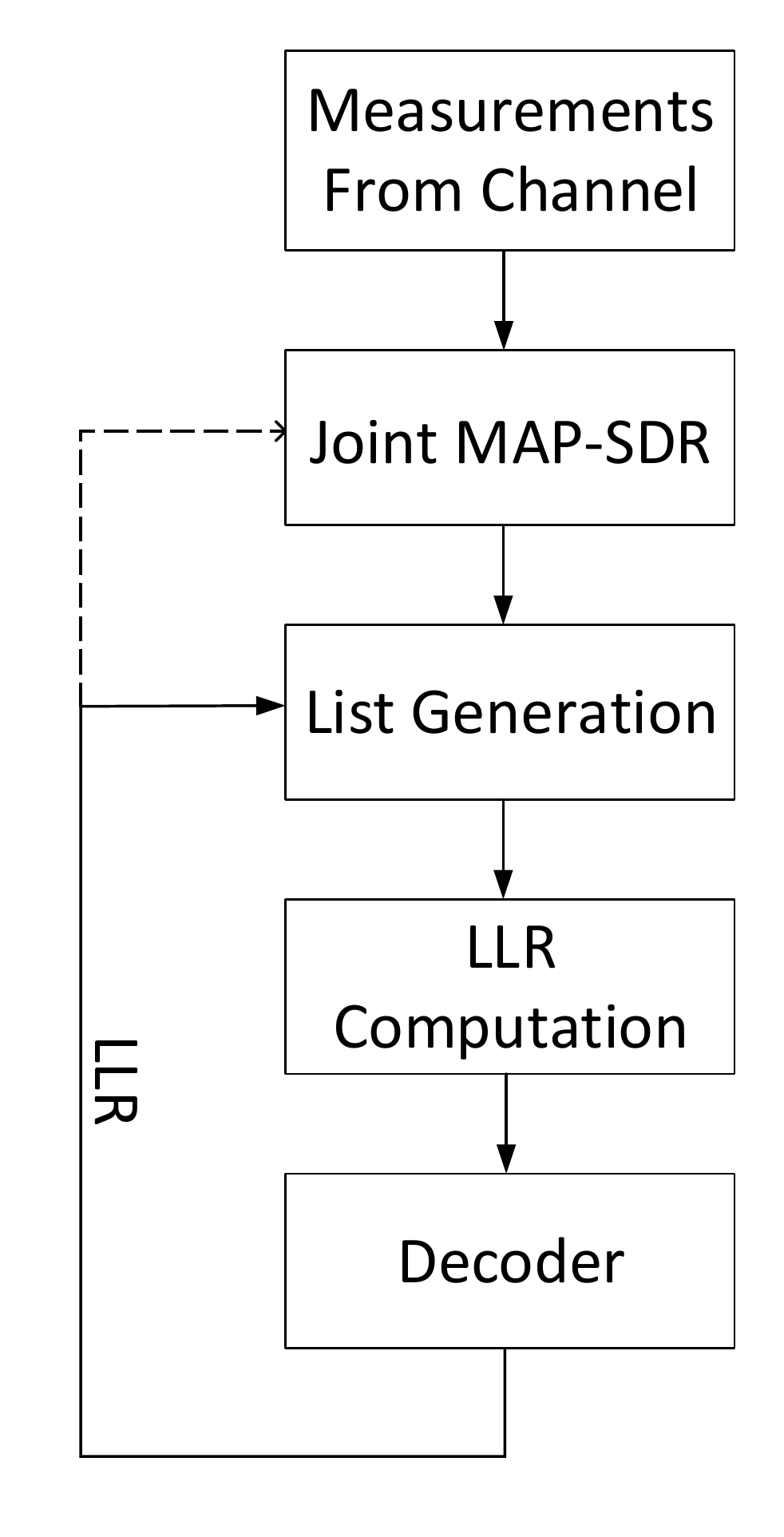}}
	      \centerline{(b)}\medskip
	    \end{minipage}
	    \caption{\small{(a) Flow of Multi Joint SDR. (b) Flow of Single Joint SDR.}}
	    \vspace*{-5mm}
	    \label{fig:flowcharts}
	\end{figure}

The underlying principle of turbo receiver is that soft detector should use information from both received signals and decoder feedback
to improve receiver performance from one iteration to another. 
During the initial iteration, we solve the joint MAP-SDR with $\mathbf{L}_{A1} = \mathbf{0}$.
The extrinsic LLR from this first iteration is denoted as $\mathbf{L}_{E1}^{init}$, 
which corresponds to the information that can be extracted from received signals.
When \textit{a priori} LLR value $\mathbf{L}_{A1}$ becomes available 
after the first iteration,
we combine them directly 
as $\mathbf{L}_{E1}^{comb} = \mathbf{L}_{E1}^{init} + \mathbf{L}_{A1}$,
and perform hard decision on $\mathbf{L}_{E1}^{comb}$ to obtain the bit vector $\mathbf{b}_k^*$ 
for each snapshot $k$, i.e., $\mathbf{b}_k^* = \text{sign}(\mathbf{L}_{E1}^{comb})$.
We then can generate list $\mathcal{\hat{L}}_k$ as before 
according to a pre-specified $P$.
The comparison with multiple-SDR turbo receiver is 
illustrated by flowcharts in Fig.~\ref{fig:flowcharts}.

We note that $\mathbf{L}_{A1}$ varies 
from iteration to iteration,
so does $\mathbf{L}_{E1}^{comb}$.
If $\mathbf{L}_{A1}$ converges towards a ``good solution'', it 
would enhance $\mathbf{L}_{E1}^{comb}$.
If $\mathbf{L}_{A1}$ is moving towards a ``poor solution'', 
then the initial LLR $\mathbf{L}_{E1}^{init}$ should help readjust
$\mathbf{L}_{E1}^{comb}$ to certain extent. 
In particular, the joint MAP-SDR detector 
(in the first iteration) can provide
a reliably good starting point $\mathbf{L}_{E1}^{init}$ for the turbo receiver,
and then additional information that can be extracted from 
resolving MAP-SDR in subsequent iterations is quite limited. 
As will be shown in our simulations, this simple receiver scheme can
generate output performance that is close to the
original algorithm that requires solving joint MAP-SDR 
in each iteration.

% ------------------------ Fig begins ------------------------
% ------------------------ Fig ends --------------------

\section{Simulation Results} \label{sec:sim}
In the simulation tests, a MIMO system with $N_t = 4$ and $N_r = 4$ is assumed.
The MIMO channel coefficients are assumed to be ergodic Rayleigh fading.
QPSK modulation is used and a regular (256,128) LDPC code with column weight 3 is employed.
We name the turbo receiver using Eq.~(\ref{eq:extr_llr}) the \textit{full list} turbo receiver. 

\subsection{Joint MAP-SDR Turbo Receiver Performance}
\begin{figure}[!tb]
\centering
\centerline{\includegraphics[width=8cm]{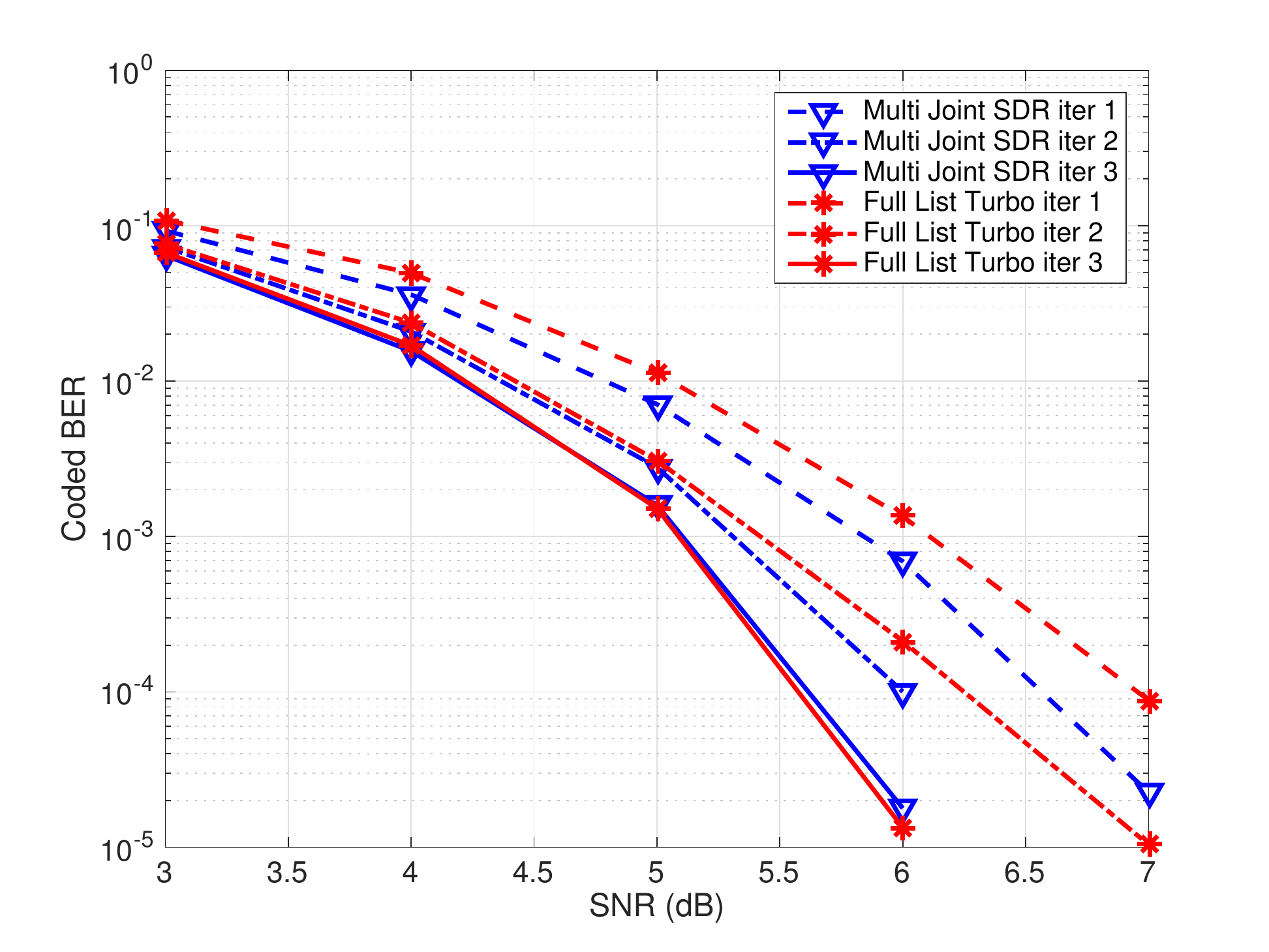}}
\caption{\small{BER comparisons of full list turbo receiver and joint MAP-SDR turbo receiver at iteration = 1, 2 and 3.}}
\label{fig:iter_comp}
\end{figure}

We investigate the performance of joint MAP-SDR turbo receiver versus full list turbo receiver.
In this test, we are more focused on the performance aspect with less concern on complexity, 
therefore we choose to run joint MAP-SDR in each iteration.
We name this turbo receiver \textit{multi joint SDR} in the figure legend.
We set Hamming radius $P=2$ and clipping value 8 for $\mathbf{L}_{E1}$.
Fig.~\ref{fig:iter_comp} shows the BER performance of 1st, 2nd and 3rd iterations. 
It is clear that joint MAP-SDR produces even better results than full list turbo in the 1st iteration. 
In later iterations, full list turbo receiver gradually catches up 
and eventually their performances become similiar. 
%\textcolor{red}{We comment that the superior performance of MAP-SDR in the 1st iteration is
% because the maximizer in subset $\mathcal{\overline{L}}_k$ could be the ``true'' maximizer
% whereas the maximizer in set $\mathcal{L}$ might not be the true one due to noise perturbation}. 

\begin{figure}[!tb]
\centering
\centerline{\includegraphics[width=9cm]{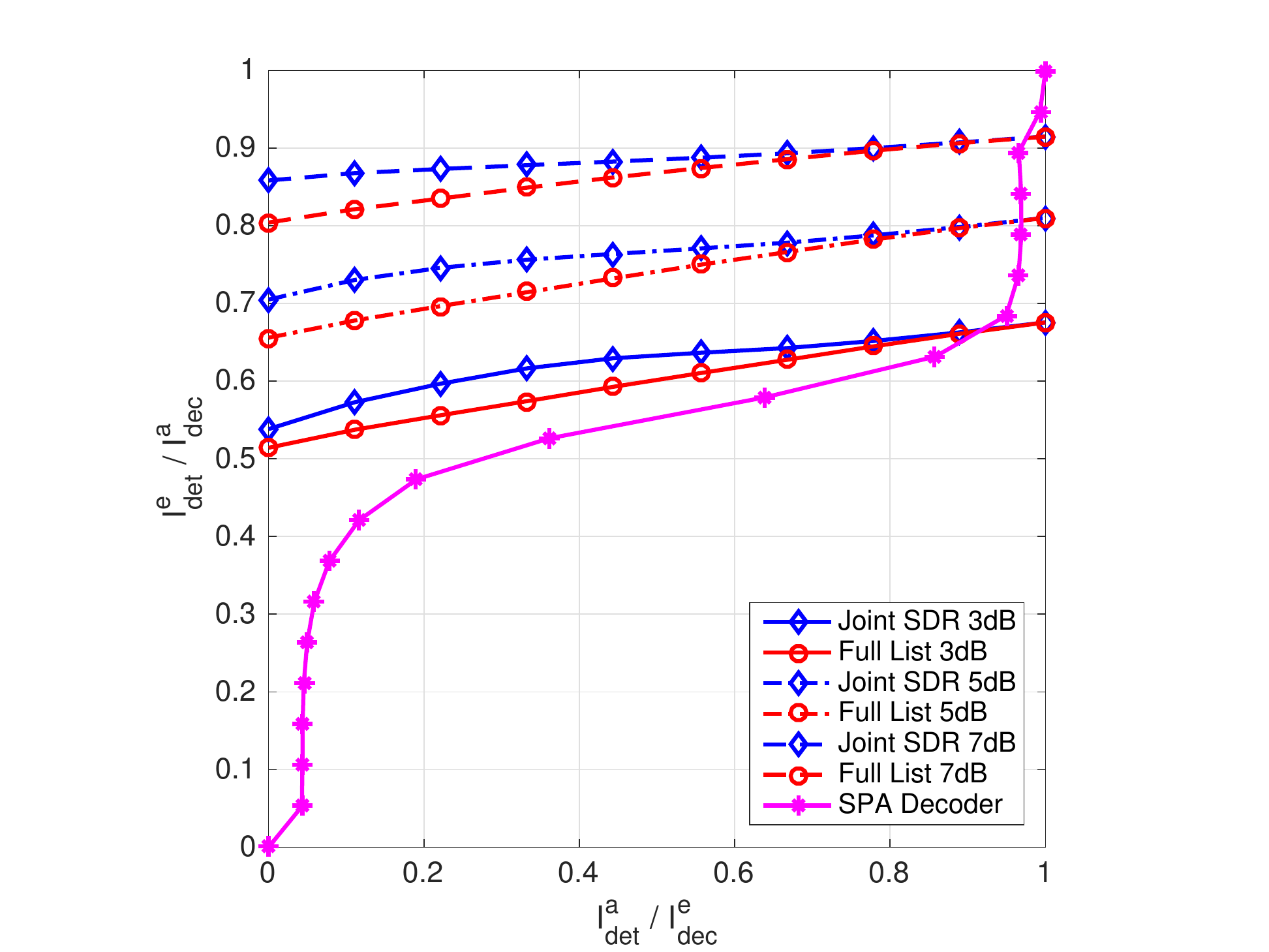}}
\caption{\small{EXIT charts of turbo equalizer and iterative SDR receiver at SNR = 3, 5 and 7 dB.}}
\label{fig:exit_comp}
\end{figure}

We also plot the extrinsic information transfer (EXIT) charts of turbo receivers that are based on joint MAP-SDR 
and full list in Fig.~\ref{fig:exit_comp} to corroborate the BER performance at various SNRs. 
Here we use the histogram method to measure the extrinsic information \cite{rob}.
When \textit{a priori} mutual information (MI) is low, the output MI of joint MAP-SDR is much higher than that of full list. 
As iteration goes, MI becomes higher, and their gap becomes smaller.

\subsection{Simplified SDR Turbo Receiver Performance}
\begin{figure}[!tb]
\centering
\centerline{\includegraphics[width=8cm]{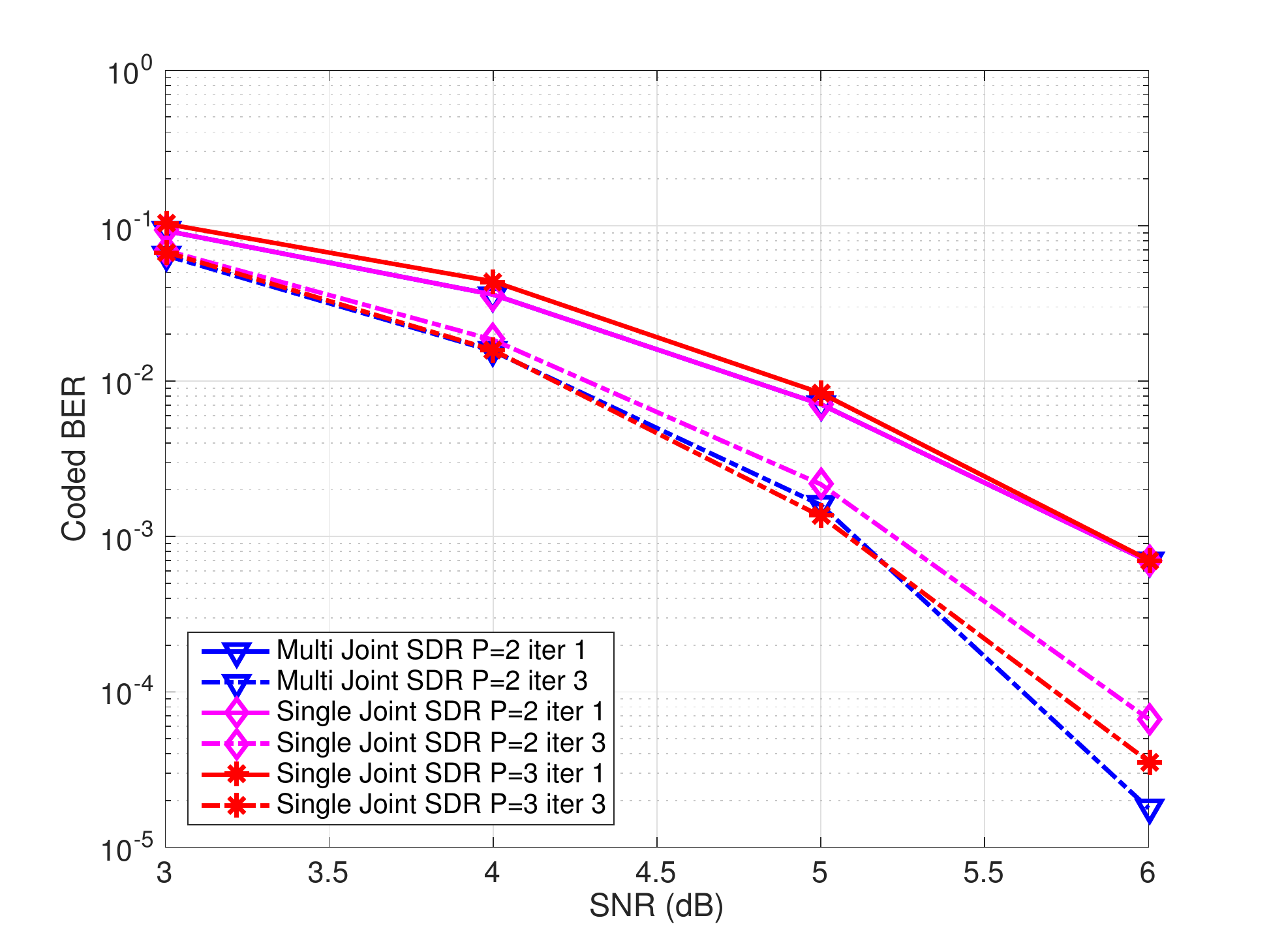}}
\caption{\small{BER comparisons of multi SDR and single SDR turbo receivers at iteration = 1 and 3.}}
\label{fig:single_sdr}
\end{figure}

The performance of \textit{single joint SDR} turbo receiver, which only runs joint MAP-SDR receiver in the initial iteration, 
is shown in Fig.~\ref{fig:single_sdr} in comparison with the \textit{multi joint SDR} that runs joint MAP-SDR in each iteration.
We choose two Hamming radii $P = 2$ and 3 for single joint SDR, while that for multi joint SDR is fixed at 2.
It is clear that they all perform equally good in the first iteration since the same joint MAP-SDR is invoked in that iteration.
At the 3rd iteration, single joint SDR slightly degrades, especially for $P=2$, but the performance degradation is acceptable 
in trade for such low complexity.

\subsection{Comparison with Other SDR Receivers}
\begin{figure}[!tb]
\centering
\centerline{\includegraphics[width=8cm]{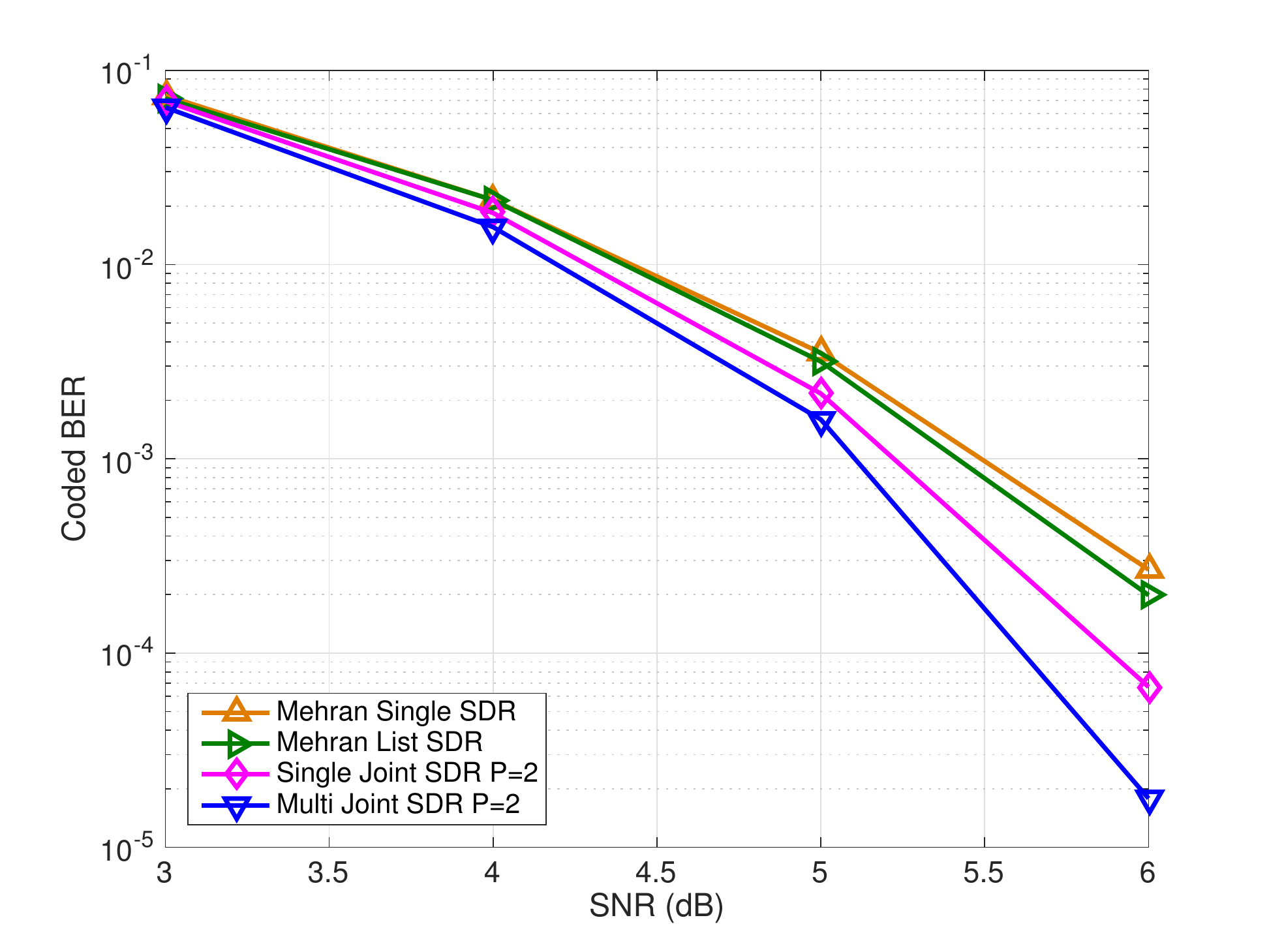}}
\caption{\small{BER comparisons of different turbo SDR receivers.}}
\label{fig:ber_comp_all}
\end{figure}

Now we compare our proposed joint SDR turbo receivers with those SDR turbo receivers from \cite{nekuii2011efficient},
which we name as ``Mehran List SDR'' and ``Mehran Single SDR'', respectively. 
The ``Mehran List SDR'' solves SDRs in each iteration while ``Mehran Single SDR''
runs one SDR in the first iteration only. 
For Mehran's methods, we employ same setting as in his paper \cite{nekuii2011efficient}: 
25 randomizations, (at most) 25 preliminary elements in the list, of which 5 elements are used for enrichment. 
All BER curves plotted in Fig.~\ref{fig:ber_comp_all} are after the 3rd iteration of turbo processing. 
For our joint SDR turbo receivers, Hamming radius $P=2$ for list generation. 
The performance advantage of our receivers is clear around BER = 1e-4. 
Both our multi SDR and single SDR receiver outperform its counterpart,
and our single SDR receiver even outperforms  ``Mehran List SDR'' that solves SDRs in each iteration. 

%\begin{figure}[!htb]
%\centering
%\centerline{\includegraphics[width=12cm]{ber_long_code_without_turbo.eps}}
%\caption{\small{BER comparisons of different Turbo SDR: (1024,512) code.}}
%\label{fig:ber_comp_long}
%\end{figure}
%
%\begin{figure}[!htb]
%\centering
%\centerline{\includegraphics[width=12cm]{time_compare_all.eps}}
%\caption{\small{Average runtime comparisons of different Turbo SDR: (256,128) code.}}
%\label{fig:time_comp_all}
%\end{figure}

\section{Conclusion} \label{sec:con}
This work presents the novel joint MAP-SDR turbo receiver and its simplified version.
The proposed receivers perform similarly to full list turbo receiver, 
while computation cost is reduced from exponential to polynomial.
Moreover, the joint SDR receivers outperform existing SDR-based turbo receivers 
by an obvious gain. 
To strengthen this work, we will conduct complexity analysis in future works.
In addition, we would like to extend the current work for higher order QAM constellations.

\bibliographystyle{IEEEtran}
\bibliography{IEEEabrv,mybibfile}
%\begin{thebibliography}{1}
%\end{thebibliography}

\end{document}